 \documentclass{article}
\usepackage{setspace}
\usepackage{dcolumn}
\usepackage{mathrsfs}
\usepackage[utf8]{inputenc}
\usepackage[margin=1in]{geometry}
 \usepackage{hyperref}
 \hypersetup{
    pdftoolbar=true,        
    pdfmenubar=true,      
    pdffitwindow=false,    
    pdfstartview={FitH},
    pdfauthor={Sabina Yeasmin},    
    colorlinks=true,      
    linkcolor=blue,        
    citecolor=blue,      
}
 \usepackage{amsmath}
 \usepackage{amssymb}
 \usepackage[backend=biber,sorting=none]{biblatex}
\usepackage{graphicx}
  \usepackage{caption}
\usepackage{subcaption}
\title{Warm Inflation in a Braneworld Scenario}
\author{Sabina Yeasmin \footnote{Email: sabina.yeasmin@aus.ac.in}, Atri Deshamukhya \\
Department of Physics, Assam University, Silchar, India \\
}
\date{}
\begin{document}

\maketitle

\begin{abstract}
 
 In the literature, many warm inflationary models are formulated. In this piece of work, a warm inflationary model in the braneworld scenario is studied, considering constant and variable dissipation coefficients. Performance of the model has been considered in both strong and weak dissipative regimes. We study the dynamics of this scenario under slow-roll approximation and estimate cosmological observables, viz., the spectral index and tensor-to-scalar ratio. In order to constrain the parameters in our model, we consider data from Planck 2018 and BICEP.
 
\end{abstract}

\section{Introduction}
\label{intro}
In the past few years, there has been much interest in studying the extra-dimensional theories of gravity as they can solve some long-standing problems in cosmology. Nordstrom \cite{r6} proposed the idea of extra dimensions in the early twentieth century, and later by Kaluza \cite{r1} and Klein \cite{r100} to unify electromagnetism and gravity by proposing a theory in five-dimensional space-time where the photon originates from the fifth component of the metric. The extra dimensions in the work of Kaluza and Klein are considered to be compact. But later N. Arkani-Hamed et al. \cite{r2} and I. Antoniadis et al. \cite{r3} suggested that the extra dimensions may be very large if our universe is confined to the three usual dimensions called the brane embedded in a higher-dimensional space called the bulk. The brane-world model is a well-studied example of the higher-dimensional theory, which has been inspired from M-theory. The first braneworld model was proposed by Randall and Sundrum (RS) in 1999 \cite{r7}, with the primary goal of solving the Hierarchy problem between the electroweak scale and the Planck scale. According to such models, the standard model of particle physics is confined to a four-dimensional brane, whereas gravity is free to access the extra dimensions. In other words, we live in a brane universe embedded in a higher-dimensional space-time called bulk. These extra dimensions (bulk) are inaccessible to ordinary matter \cite{r2} but contribute to the Einstein field equations on the brane. This contribution introduces some modifications in the fundamental field equations, such as the Friedmann equation. Such corrections in the Friedmann equation significantly impact the inflationary scenario in the early universe.

 The inflationary scenario is a successful paradigm describing the evolution of the universe at a very early era. It is an epoch in the early universe in which the scale factor was thought to be expanded exponentially and the idea is capable of providing answers to some of the major issues associated with the standard Big Bang Model \cite{r30}. In addition, it provides the mechanism for generating primordial density fluctuations that are believed to provide initial seeds of the CMB anisotropies, leading to the formation of large-scale structures we observe in the universe today. The experimental data from cosmic microwave background radiation (CMBR), large-scale structure (LSS), cosmic background explorer (COBE), WMAP, Planck satellite, etc, provide evidence in favor of such an idea \cite{r36, r37, r38, r39}. In standard inflationary models, the expansion of the universe is driven by a scalar field called inflaton, which does not interact with any other field during the inflationary phase \cite{r31,r32,r33,r34,r35}. So, particle production was not possible during inflation, leading the universe to a supercooled phase. To shift from a supercooled phase to a radiation-dominated big-bang phase, a process called the reheating phase is required. During the reheating phase, the inflaton oscillates around the minimum of its potential. This oscillation would lead to the production of particles and reheat the universe.

Warm inflation is the alternative dynamical realization of inflation in which the inflaton field is assumed to be coupled to other fields and decays its vacuum energy into the radiation field during the inflationary period \cite{r40,r41,r42}. This means that the universe is not supercooled during inflation, but instead, there is a continuous production of radiation that keeps the universe at a finite temperature $\tau>H$ without the requirement of any separate reheating phase \cite{r21,r22}. The interaction of the inflaton field and radiation field leads to dissipative effects that modify the warm inflationary dynamics. Thus, a term $\Gamma$ called dissipation coefficient is added to the dynamical equations of warm inflation which describes the dissipation of the inflaton field energy into the thermal bath. The idea of warm inflation was first introduced by Berera in 1995 \cite{r40}, and after that several aspects of warm inflationary scenarios have been addressed in literature \cite{r43,r44,r45,r46,r47}. Over the past two decades, numerous warm inflation models have been extensively studied in the braneworld scenario \cite{r48,r4,r50,r51,r52,r53} with different potentials. Considering different forms of dissipation coefficients, they have studied the impact of dissipation parameters in generating cosmological perturbations in their model on the brane. In this work, we study a warm inflationary model with natural potential in the context of braneworld cosmology, considering a constant and a temperature-dependent dissipation coefficient. 

This paper is organized as follows: In Section 2, we briefly discuss the basic dynamics of warm inflation in the braneworld scenario and investigate the cosmological perturbations in our model in Section 2.1. In Section 2.2, we briefly address the dynamics of axion-like particles, and in Sections 2.3 and 2.4, we analyse the model numerically in the strong and weak dissipation regimes, considering both a constant and a temperature-dependent dissipation coefficient. And, finally, in Section 3, we give our conclusions.

 \section{Warm inflation dynamics in the braneworld scenario }
  We consider a braneworld scenario in which the matter fields are confined to the brane embedded in a five-dimensional bulk. The total action that contains Einstein-Hilbert action and brane action is given by \cite{r54},
\begin{eqnarray}
   S=\int d^5x\sqrt{-g^{(5)}} \left(\frac{M_5^3}{2}R+L_{bulk}\right)+\int d^4x\sqrt{-g} \left(M_5^3 K^{\pm}+L_{brane}\right)
\end{eqnarray}
 where, $M_5$ is the 5-dimensional Planck mass, $R$ is the five-dimensional Ricci scalar, $L_{bulk}$ is the matter Lagrangian in the bulk, $K^{\pm}$ is the trace of the extrinsic curvature on either side of the brane, $L_{brane}$ is the Lagrangian of matter confined on the brane. By varying the above action with respect to the bulk metric, the 5-D Einstein equations in the bulk are obtained as \cite{r54,r55}
\begin{eqnarray}
    G_{AB}^{(5)}=-\Lambda_5 g_{AB}^{(5)}+\frac{1}{M_5^3} T_{AB}^{(5)}
\end{eqnarray}
Here, $\Lambda_5$ represents the bulk cosmological constant and $T_{AB}^{(5)}$ represents the 5-D energy-momentum tensor of bulk matter. The basic equations in the brane-world scenario are derived by projecting the higher-dimensional variables onto the brane. The induced metric on brane is $g_{AB}=g_{AB}^{(5)}-n_A n_B$ with $n^A$ as the unit vector normal to brane hypersurface \cite{r54}. Now, the Gauss equation, which describes the projection of the 5-D curvature on the brane, is given by \cite{r54,r55,r16}  
\begin{eqnarray}
R_{ABCD}^{(4)}=R_{EFGH}^{(5)}g_{A}^Eg_{B}^Fg_{C}^Gg_{D}^H+K_{AC}K_{BD}-K_{AD}K_{BC}
\label{ee3}
\end{eqnarray}
and the Codazzi equation, which determines the change of extrinsic curvature of the brane is given by
\begin{eqnarray}
    R_{BC}^{(5)}g_{A}^B n^{C}=\nabla_B K_A^B-\nabla_A K
\end{eqnarray}
where $K_{AB}=g_A^M g_B^N \nabla_M n_N$ is the extrinsic curvature of the brane. Contracting the Gauss equation (\ref{ee3}), effective 4-D equations can be obtained as
\begin{eqnarray}
\label{ee5}
G_{\alpha\beta} &=& -\frac{1}{2} \Lambda_5 g_{\alpha\beta} + \frac{2}{3 M_5^3}  
\left[T_{AB}^{(5)} g^A_{\ \alpha} g^B_{\ \beta} + 
\left( T_{AB}^{(5)} n^A n^B - \frac{1}{4} T^{(5)} \right) g_{\alpha\beta} \right]
\\ \nonumber &+& K K_{\alpha\beta} - K_{\alpha}^{\ \mu} K_{\mu\beta}  + \frac{1}{2}
\left( K^{\mu\nu} K_{\mu\nu} - K^2 \right) g_{\alpha\beta} - E_{\alpha\beta}
\end{eqnarray}
where
\begin{equation}
E_{\alpha\beta} =  C_{ABCD}^{(5)}\, n^C n^D\, g^A_{\ \alpha} g^B_{\ \beta},
\label{ee6}
\end{equation}
is a projection of the five-dimensional Weyl tensor. 

The total energy-momentum tensor on the brane is
\begin{equation}
    T_{\alpha\beta}^{brane}=T_{\alpha\beta}-\lambda g_{\alpha\beta}
\end{equation}
where $T_{\alpha\beta}$ is the energy–momentum tensor of matter fields confined to the brane and $\lambda$ is the brane tension. The 5-D field equations, including the brane world contributions, has the form
\begin{eqnarray}
    G_{AB}^{(5)}=-\Lambda_5 g_{AB}^{(5)}+\frac{1}{M_5^3}\left[T_{AB}^{(5)}+T_{AB}^{brane}\delta(y)\right]
\end{eqnarray}
Here, $\delta(y)$ is the localization of brane contributions. The singular behavior of the energy–momentum tensor of the brane leads to the Israel junction conditions as
\begin{equation}
    g^+_{\alpha\beta}-g^-_{\alpha\beta}=0
\end{equation}
\begin{equation}
    K^+_{\alpha\beta}-K^-_{\alpha\beta}=-\frac{1}{M_5^3}\left[T_{\alpha\beta}^{brane}-\frac{1}{3}T^{brane}g_{\alpha\beta}\right]
    \label{ee10}
\end{equation}
Substituting equation (\ref{ee10}) into equation (\ref{ee5}), the projection of the Einstein field equations onto the brane is obtained, given by \cite{r54,r55,r16},
\begin{equation}
    G_{\alpha\beta}=-\Lambda_4 g_{\alpha\beta}+\left(\frac{8\pi}{M_4^2}\right)T_{\alpha\beta}+\left(\frac{8\pi}{M_5^3}\right)^2 \pi_{\alpha\beta}-E_{\alpha\beta}
    \label{e3}
\end{equation}
$E_{\alpha\beta}$ describes the effect of the bulk graviton on the brane, $T_{\alpha\beta}$ is the stress-energy tensor on the brane, $M_5$ is the Planck mass in $5$ dimensions, $\pi_{\alpha\beta}$ is a tensor quadratic in $T_{\alpha\beta}$ given by
\begin{equation}
    \pi_{\alpha\beta}=-\frac{1}{4}T_{\alpha\beta} T_\beta^\beta+\frac{1}{12}T T_{\alpha\beta}+\frac{1}{24}(3T_{\beta\gamma}T^{\beta\gamma}-T^2)g_{\alpha\beta}
    \label{e4}
\end{equation}

% In this section, we consider a 5-D brane cosmology in which the matter fields are confined to the brane embedded in a five-dimensional bulk. The projection of Einstein field equations onto the brane with a cosmological constant is given by \cite{r16},
% \begin{equation}
%     G_{\alpha\beta}=-\Lambda_4 g_{\alpha\beta}+\left(\frac{8\pi}{M_4^2}\right)T_{\alpha\beta}+\left(\frac{8\pi}{M_5^3}\right)^2 \pi_{\alpha\beta}-E_{\alpha\beta}
%     \label{e3}
% \end{equation}
% where $E_{\alpha\beta}$ is a projection of the five dimensional Weyl tensor and it describes the effect of the bulk graviton on the brane, $T_{\alpha\beta}$ is the stress-energy tensor on the brane, $M_5$ is the five-dimensional Planck mass, $\pi_{\alpha\beta}$ is a tensor quadratic in $T_{\alpha\beta}$ 
% \begin{equation}
%     \pi_{\alpha\beta}=-\frac{1}{4}T_{\alpha\nu} T_\beta^\nu+\frac{1}{12}T T_{\alpha\beta}+\frac{1}{24}(3T_{\nu\gamma}T^{\nu\gamma}-T^2)g_{\alpha\beta}
%     \label{e4}
% \end{equation}
 
The effective cosmological constant $\Lambda_4$ on the brane is related to the five-dimensional cosmological constant $\Lambda$ and the brane tension $\lambda$ as
\begin{equation}
\Lambda_4=\frac{4\pi}{M_5^3}\left(\Lambda+\frac{4\pi}{3M_5^3}\lambda^2\right)
    \label{e5}
\end{equation}
Four-dimensional Planck mass $M_4$ is related to the five-dimensional Planck mass $M_5$ as
\begin{equation}
M_4=\sqrt{\frac{3}{4\pi}}\left(\frac{M_5^2}{\sqrt{\lambda}}\right)M_5
    \label{e6}
\end{equation}
The projection of the Einstein equations onto the brane modifies the Friedmann equation. The modified Friedmann equation on this braneworld scenario has the form \cite{r56,r17,r18}:
\begin{equation}
    H^2= \frac{\Lambda_4}{3}+\left(\frac{8\pi}{3M_4^2}\right)\rho+\left(\frac{4\pi}{3M_5^3}\right)^2\rho^2+\frac{\varepsilon}{a^4}
    \label{e7}
 \end{equation}
where $\rho$ is the total energy density of the matter content of the universe on the brane, $\varepsilon$ is an integration constant arising from $E_{\alpha\beta}$. The last term $\frac{\varepsilon}{a^4}$ will be rapidly diluted once inflation begins, and we can neglect it over most of the inflationary epoch. $\Lambda_4$ is assumed to be negligible in the early universe. This is achieved by imposing the fine-tuning $\Lambda=-\frac{4\pi\lambda^2}{3M_5^3}$ , which means that the bulk cosmological constant compensates the brane tension, leading to an effectively vanishing $\Lambda_4$ on the brane. With these assumptions, Friedmann equation (\ref{e7}) reduces to

\begin{equation}
    H^2=  \frac{8\pi}{3M_4^2}\rho\left(1+\frac{\rho}{2\lambda}\right)
    \label{e8}
\end{equation} 
Thus, the presence of a brane introduces a term $\frac{8\pi \rho^2}{6M_4^2\lambda}$ in the classical Friedmann equation. In the high-energy scale, i.e. when $\rho\gg \lambda $, such a term will be dominant, which makes the early universe cosmology different from the one described by the standard scenario. In the low-energy scale, i.e, when $\rho\ll \lambda $, the results of general relativity is recovered.
%In the warm inflation scenario, the universe is filled by a homogeneous scalar field called inflaton $\phi$ with energy density $\rho_\phi$ and radiation field with energy density $\rho_\gamma$. In our model, we consider an axion-like particle as the inflaton.
In the considered setup, axion-like particles propagate in the five–dimensional bulk 
% at finite temperature
and get confined to the brane after dimensional reduction. 
% Since the temperature of such a universe is non-zero, 
The total energy density confined on the brane is given by,
\begin{equation}
    \rho=\rho_\phi+\rho_\gamma=\frac{\dot\phi^2}{2} + V(\phi)  +\rho_\gamma
    \label{e9}
\end{equation}
where $V(\phi)$ is the potential of the axionic field, which we consider as the field driving inflation.

The following set of equations describes the background dynamics of this model:
\begin{equation}
    H^2=  \frac{8\pi}{3M_4^2}\left(\frac{\dot\phi^2}{2} + V(\phi)  +\rho_\gamma\right)\left(1+\frac{\frac{\dot\phi^2}{2} + V(\phi)  +\rho_\gamma}{2\lambda}\right)
    \label{e10}
\end{equation}
\begin{equation}
\dot{\rho}_\phi+3H(\rho_\phi+P_\phi)=-\Gamma \dot{\phi}^2
\label{e11}
\end{equation}
\begin{equation}
    \dot{\rho}_\gamma+4H\rho_\gamma=\Gamma \dot{\phi}^2
    \label{e12}
\end{equation}  
where $P_\phi$ is the pressure of the scalar field given by $P_\phi=\frac{\dot\phi^2}{2}-V(\phi) $. The equation of motion of the inflaton field can be derived by substituting the form of $\rho_\phi$ and $P_\phi$ into equation \eqref{e11}, 
\begin{eqnarray}
  \ddot \phi+3H\dot \phi (1+Q)+V^\prime = 0
  \label{e13}
\end{eqnarray}
Where the prime denotes the derivative with respect to the inflaton field, $Q$ is the dissipation rate, defined as
\begin{equation}
     Q=\frac{\Gamma}{3H}
     \label{e14}
\end{equation}
The dissipation rate $Q$ represents the rate at which the scalar field decays into radiation during inflationary expansion. Depending on the values of $Q$, there can be two separate dissipation regimes in warm inflation, the strong ($Q\gg1$) and weak ($Q\ll1$) dissipative regimes \cite{r19,r20}.  We studied this model in both the dissipative regime and in high-energy scenario. The dissipation coefficient $\Gamma$ may be a constant or a function of the scalar field or the temperature of the thermal bath or both. 
% As mentioned in the introduction, here we have considered two forms of $\Gamma$, one is constant and the other one is a function of temperature $\tau$ only. 

During warm inflation, the potential energy of the inflaton field dominates over both the kinetic energy of the inflaton field and the energy density of the radiation field.  Although the radiation energy density produced during warm inflation is sub-dominant, it satisfies the condition $\rho_\gamma^{1/4}>H$, which means that the thermalization condition is $\tau>H$ \cite{r21,r22}. Also, the radiation bath produced during warm inflation is assumed to be nearly in thermal equilibrium. This approximation is known as the slow-roll approximation. These can quantitatively be expressed as 
 \begin{equation}
     \frac{ 1}{2}\dot{\phi}^2+\rho_\gamma<< V(\phi)
     \label{e15}
\end{equation}
 \begin{equation}
 \ddot \phi << 3H\dot \phi (1+Q) 
 \label{e16}
 \end{equation}
 \begin{equation}
    \dot{\rho}_\gamma<< 4H\rho_\gamma 
    \label{e17}
\end{equation}

Under the slow-roll approximations, warm inflation equations (\ref{e10}), (\ref{e13}) and (\ref{e12}) read as
 \begin{equation}
    H^2=  \frac{8\pi}{3M_4^2}  V(\phi)   \left(1+\frac{ V(\phi)}{2\lambda}\right)
    \label{e18}
\end{equation}
 \begin{equation}
       3H\dot \phi (1+ Q) + V^\prime = 0
       \label{e19}
 \end{equation}
 \begin{equation}
     \rho_\gamma=\frac{\Gamma \dot{\phi}^2}{4H} =C\tau^4
     \label{e20}
 \end{equation}
  Where $\tau$ is the temperature of the radiation bath, $C=\frac{g_*\pi^2}{30}$ is the Stefan-Boltzmann constant, and $g_*$ is the effective number of degrees of freedom for the radiation at temperature $\tau$ ( we take $ g_*= 200$).
 
 The slow-roll approximation can be parameterized by a set of slow-roll parameters defined as
\begin{equation}
    \epsilon=-\frac{\dot H}{H^2}=\frac{M_4^2}{16\pi}\left(\frac{V^\prime}{V}\right)^2\frac{4\lambda(\lambda+V)}{(2\lambda+V)^2}\frac{1}{(1+Q)}
    \label{8}
\end{equation}
\begin{equation}
    \eta=-\frac{\ddot H}{\dot H H}=\frac{M_4^2}{8\pi}\left(\frac{V^{\prime\prime}}{V}\right)\frac{2\lambda}{(2\lambda+V)}\frac{1}{(1+Q)}
    \label{9}
\end{equation}
In the low-energy ($V\ll\lambda$) and weak dissipative ($Q\ll1$) regime, the above slow-roll parameters reduce to those in the standard inflationary model. The slow-roll parameters have to satisfy the conditions $\epsilon<<1$ and $|\eta|<<1$  during the inflationary phase, and the violation of these conditions marks the end of inflation.

The amount of expansion of the universe throughout the inflationary epoch is measured by the number of e-folds. For the model, the number of e-foldings when the inflation field $\phi$ rolls from its value $\phi_i$ to $\phi_f$ is given by

 \begin{eqnarray}
N&=&\int^{t_2}_{t_1} H dt=\int^{\phi_f}_{\phi_i}\frac{H}{\dot{\phi}}d\phi\nonumber\\ &=& - \frac{8\pi}{M_4^2}\int _{\phi_i}^{\phi_f}\frac{V}{V^\prime}\left(1+\frac{V}{2\lambda}\right)(1+Q)d \phi
\label{115}
\end{eqnarray}

\subsection{Perturbation calculation}
  In warm inflation, both the background inflationary dynamics and cosmological perturbations get modified due to the presence of dissipative effects. These effects generate a thermal bath during the inflationary phase. Thus, in warm inflation, inflaton fluctuations are sourced by thermal noise, unlike in cold inflation where they arise from quantum fluctuations. The primordial power spectrum in warm inflation can be expressed as \cite{r23,r24,r25,r26}
\begin{eqnarray}
    P_{R}&=&\frac{H^2}{\dot{\phi}^2}\left[ \frac{(\Gamma+H)\tau}{16\pi^2}\frac{8\pi}{\sqrt{9+12\pi Q}}\right]G(Q)
    \label{2}
\end{eqnarray}
where the term $G(Q)$ accounts for the effect of the coupling between the inflaton fluctuations and the radiation fluctuations. $G(Q)$ can be determined by numerically solving the complete set of coupled perturbation equations in warm inflation and fitting it to an appropriate function. Using WI2easy \cite{r5}, we find a functional form for $G(Q)$ for our warm inflationary model on brane, which is given by
\begin{eqnarray}
   G(Q)= \frac{0.72 e^{20.44 Q^{0.028315}}}{1+0.0394 e^{\left(0.6 Q^{0.86}+22.6686*Q^{0.001315}\right)}}+\frac{1+0.00821 Q^{1.8315}}{\left(1+0.41 Q^{0.815}\right)^{3.5}}
\end{eqnarray}
The power spectrum for the tensor perturbation is \cite{r27}
\begin{eqnarray}
P_T&=&\frac{16 \pi}{M_4^2}\left(\frac{H}{2\pi}\right)^2 \coth \left(\frac{k}{2 \tau }\right)\nonumber \\
&=&  \frac{64 \pi V^2}{9 M_4^2 M_5^6}\coth \left(\frac{k}{2 \tau }\right)  
\label{118}
\end{eqnarray}

The spectral index $n_s$ and tensor to scalar ratio $r$ are defined as \cite{r28}
\begin{equation}
  n_s-1=\frac{d \ln P_R}{d \ln k}
  \label{e16}
\end{equation}
\begin{equation}
    r=\frac{ P_T}{P_R}
    \label{e17}
\end{equation}
\subsection{Dynamics of axion-like particles}
% The motivation for considering a natural potential is that the flatness of the potential required for successful inflation is protected by the axionic shift symmetry $\phi\rightarrow \phi+$constant, where $\phi$ is the axion which plays the role of inflaton field. As long as the inflaton potential is fully flat, the inflaton can not roll down the potential and it is not possible to derive inflation. So, shift symmetry can not be exact and there must be an explicit symmetry breaking that generates the axion potential of the form
Axion-like particles are pseudo–Nambu–Goldstone bosons (PNGB) that arise when a global symmetry is spontaneously broken at some scale $f$, and then explicitly broken at a lower scale $\mu$. The potential of an axion-like particle is flat due to the axionic shift symmetry $\phi\rightarrow \phi+$constant and has the form
\cite{r11,r12,r13}
\begin{equation}
  V(\phi)= \mu^4\left(1+\cos\left(\frac{\phi}{f}\right)\right)  
  \label{e1}
\end{equation}
where $f$ is the axion decay constant and $\mu$ is the mass scale of the axion. This potential can be flat for higher values of $f$. Several natural inflationary models have been studied in standard cosmology and brane-world cosmology which are consistent with the current observational data for higher values of $f$ like $f>0.6 M_P$ \cite{r29}, where $M_P$ is the reduced Planck mass. However, such higher values of $f$ could be problematic and difficult to embed in string theory. In this study, we show that this model agrees with observational data for the axion decay constant that lies well below the Planck scale. We assumed that the axion interacts with light gauge fields through the Lagrangian term,
$\mathcal{L}_{int}\propto \frac{\phi}{f} \mathrm{Tr} F \tilde{F}$ where, $F$ is the gauge field and $\tilde{F}$ is its duel. This interaction leads to the dissipation coefficient of the  form \cite{r8}
\begin{equation}
    \Gamma(\tau)=C_\Gamma\frac{\tau^3}{f^{2}}
     \label{e2}
\end{equation}
where $\tau$ is the temperature of the thermal bath and $C_\Gamma$ is a dimensionless factor proportional to the coupling constant between the inflaton field $\phi$ and gauge field. This form of dissipation coefficient has been studied extensively in the literature \cite{r8,r9,r10}. Additionally, we also consider a constant dissipation coefficient $\Gamma_0$ and study the impact of these dissipation coefficients on the linear theory of the cosmological perturbations and the cosmological observables like spectral index and tensor-to-scalar ratio appeared in our warm inflation model on brane. We constrain these cosmological parameters using the Planck 2018 \cite{r14} and BICEP \cite{r15} data.

\subsection{Numerical results in high energy ($V\gg \lambda$) and weak dissipation ($Q\ll 1$) regime}
The slow-roll parameters in this regime are 
\begin{eqnarray}
    \epsilon=\frac{3 M_5^6}{16 \pi ^2 f^2 \mu ^4}\frac { \left(1-\cos \left(\frac{\phi }{f}\right)\right)}{  \left(\cos \left(\frac{\phi }{f}\right)+1\right)^2}
\end{eqnarray}
\begin{eqnarray}
   \eta= -\frac{3 M_5^6 \cos \left(\frac{\phi }{f}\right)}{16 \pi ^2 f^2 \mu ^4 \left(\cos \left(\frac{\phi }{f}\right)+1\right)^2}
\end{eqnarray}
 Inflation ends when either of the slow-roll conditions is violated. Numerically, it is found that the parameter $\epsilon$ is the first to violate the slow-roll approximation. Hence, the field value at the end of inflation, $\phi_f$ is determined from the condition $\epsilon(\phi_{f})=1$ and is given by
 \begin{eqnarray}
     \phi_f= f\left(-\cos ^{-1}\left(\frac{-32 \pi ^2 f^2 \mu ^4+\sqrt{3} \sqrt{128 \pi ^2 f^2 \mu ^4 M_5^6+3 M_5^{12}}-3 M_5^6}{32 \pi ^2 f^2 \mu ^4}\right)\right)
 \end{eqnarray}
 The number of e-foldings (equation \eqref{115}) in this regime has the form, 
  \begin{eqnarray}
N =- \frac{16\pi^2}{3M_5^6}\int _{\phi_i}^{\phi_f}\frac{V^2}{V^\prime}d \phi
\label{ee115}
\end{eqnarray}
Solving equation \eqref{ee115} for N=60, the initial value of the inflaton field $(\phi_i)$ is obtained. Since, in the weak dissipation regime, Hubble friction dominates the particle production friction, so, the coupling between the radiation and the inflaton fields can be neglected. However, the presence of a thermal bath can still modify the expression of the primordial power spectrum as
\begin{eqnarray}
    P_R=\frac{H^3\tau}{2\pi^2\dot\phi^2}
\end{eqnarray}

\subsubsection{Case-1: $\Gamma=\Gamma_0$}
The amplitude of density perturbations is highest when evaluated at the initial field value $\phi_i$. In this case, the scalar power spectrum $P_R$, the tensor power spectrum $P_T$, and the spectral index $n_s$ at $\phi_i$ can be evaluated as,
\begin{equation}
    P _R= \frac{128 \pi ^{9/4} f^{3/2} {\Gamma_0}^{1/4}   \left(\mu ^4 \left(\cos \left(\frac{\phi_i }{f}\right)+1\right)\right)^{17/4}}{\ 3^{11/4} {C}^{1/4} M_5^{51/4} \mu ^6 \sin ^{\frac{3}{2}}\left(\frac{\phi_i }{f}\right)}
\end{equation}
\begin{equation}
    P_T=\frac{64\pi  \left( \mu^4\left(1+\cos\left(\frac{\phi_i}{f}\right)\right)  \right)^2 \coth \left(\frac{k}{2 \tau  }\right)}{9 M_4^2 M_5^6}
\end{equation}
\begin{equation}
   n_s= 1-\left(\frac{51 M_5^3 \left(\mu ^4 \sin \left(\frac{\phi_i}{f}\right)\right)^2}{16 \pi f^2    \left( \mu^4\left(1+\cos\left(\frac{\phi_i}{f}\right)\right)  \right)^2}+\frac{9 M_5^3 \left(\mu ^4 \cos \left(\frac{ \phi_i}{f}\right)\right)}{  8 \pi f^2     \mu^4\left(1+\cos\left(\frac{\phi_i}{f}\right)\right)  }\right)\frac{M_5^3}{4\pi \mu^4\left(1+\cos\left(\frac{\phi_i}{f}\right)\right)   }
\end{equation}
In order to check the viability of our model, it is necessary to compare our findings with current observational data. We predict the values of the spectral tilt and tensor-to-scalar ratio for different values of dissipation coefficients $\Gamma_0$ while keeping the parameters fixed as $\mu=0.0001 M_4$, axion decay constant $f=0.02M_4$, $C=70$ and $M_5=7.5\times 10^{-4}M_4$ for 60 e-foldings. The results are summarized in table 1. The temperature of the thermal bath $(\tau)$ and dissipation parameter $Q$ calculated for each case are also presented in the table. From the table, it is seen that in the weak dissipative regime, the spectral index does not depend on $\Gamma_0$. The range of $\Gamma_0$ (in units of $M_4$) is found to be $[9\times10^{-8}, 6\times10^{-7}]$. Within this range, the warm natural inflationary scenario on the brane yields $n_s$ and $r$ that are well within the bounds set by Planck 2018 and BICEP3/Keck Array (BK18) data. Thus, the model remains viable in the weak dissipative regime.   

\begin{table}[h!]
     
\caption{The values of the spectral index $n_s$, tensor to scalar ratio $r$, $Q$ and temperature of the thermal bath $\tau$ (in units of $M_4$) for different values of $\Gamma_0$ (in units of $M_4$) with $f=0.02M_4$, $\mu=0.0001 M_4$, $M_5=0.00075 M_4$ and $N=60$}
   
  \begin{center}
\begin{tabular}{lllll}
\hline\noalign{\smallskip}
 $\Gamma_0$ & $Q$  & $ n_s $  & $ r$   & $ \tau$ \\
 \noalign{\smallskip}\hline\noalign{\smallskip}
 % $4\times10^{-8}$ & $0.0069$ & $0.962729$ & $1.91\times10^{-7}$ & $1.58\times10^{-6}$  \\
  
 $9\times10^{-8}$ & $0.015$ & $0.962729$ & $1.906\times10^{-7}$ & $1.928\times10^{-6}$  \\
  
 $4\times10^{-7}$ & $0.069$ & $0.962729$ & $1.902\times10^{-7}$ & $2.8\times10^{-6}$ \\
 
$6\times10^{-7}$ & $0.1$ & $0.962729$ & $1.9\times10^{-7}$ & $3.12\times10^{-6}$ \\
 
% $1.2\times10^{-6}$ & $0.2$ & $0.962729$ & $1.899\times10^{-7}$ & $3.71\times10^{-6}$ \\
 
% $3\times10^{-6}$ & $0.52$ & $0.962729$ & $1.89\times10^{-7}$ & $4.66\times10^{-6}$ \\
\noalign{\smallskip}\hline

\end{tabular}
\end{center}
\label{ta1}
\end{table}

  \begin{figure}[h!]
 \centering
 \begin{subfigure}[h]{0.49\textwidth}
     \centering
     \includegraphics[width=\textwidth]{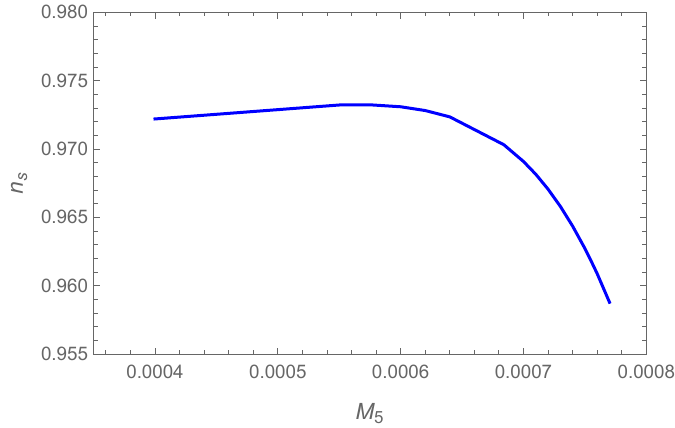}
    
 \end{subfigure}
 \hfill
 \begin{subfigure}[h]{0.49\textwidth}
     \centering
     \includegraphics[width=\textwidth]{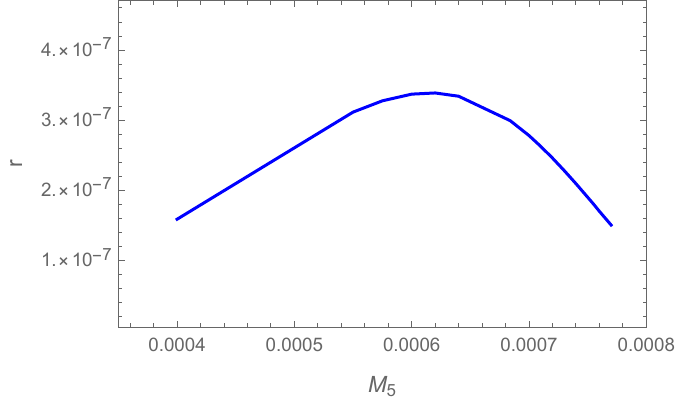}
     
 \end{subfigure}
 \caption{  Spectral index $n_s$ and tensor-to-scalar  ratio $r$ as a function of $M_5$ (in units of $M_4$) for $\Gamma_0=4\times 10^{-7} M_4$.}
    \label{f1}
 \end{figure}

Fig.~\ref{f1} shows the variation of spectral index $n_s$ and tensor-to-scalar ratio $r$ as a function of $M_5$ (in units of $M_4$) for $\Gamma_0=4\times 10^{-7} M_4$. From this plot, the range of $M_5$ (in units of $M_4$) is found to be $0.0007<M_5<0.00076$ for $\Gamma_0=4\times 10^{-7} M_4$.

\subsubsection{Case-2: $\Gamma(T)=C_\Gamma\frac{\tau^3}{f^{2}}$}In this case, the scalar power spectrum $P_R$, the tensor power spectrum $P_T$, and the spectral index $n_s$ at $\phi_i$ can be obtained as,
\begin{equation}
    P _R=\frac{2 f^{3/2} \left(\frac{1}{M_5^3}\right)^{17/4} \left(\mu ^4 \left(\cos \left(\frac{\phi_i}{f}\right)+1\right)\right)^{17/4} \sqrt[4]{\frac{C_\Gamma^4 M_5^{27} \sin ^6\left(\frac{\phi_i}{f}\right)}{C^3 f^{14} \mu ^{12} \left(\cos \left(\frac{\phi_i}{f}\right)+1\right)^9}}}{9 \sqrt[4]{C} \mu ^6 \sin ^{\frac{3}{2}}\left(\frac{\phi_i}{f}\right)}
\end{equation}
\begin{equation}
    P_T=\frac{64\pi  \left( \mu^4\left(1+\cos\left(\frac{\phi_i}{f}\right)\right)  \right)^2 \coth \left(\frac{k}{2 \tau  }\right)}{9 M_4^2 M_5^6}
\end{equation}
\begin{equation}
   n_s= 1-\left(\frac{51 M_5^3 \left(\mu ^4 \sin \left(\frac{\phi_i}{f}\right)\right)^2}{16 \pi f^2    \left( \mu^4\left(1+\cos\left(\frac{\phi_i}{f}\right)\right)  \right)^2}+\frac{9 M_5^3 \left(\mu ^4 \cos \left(\frac{ \phi_i}{f}\right)\right)}{  8 \pi f^2     \mu^4\left(1+\cos\left(\frac{\phi_i}{f}\right)\right)  }\right)\frac{M_5^3}{4\pi \mu^4\left(1+\cos\left(\frac{\phi_i}{f}\right)\right)   }
\end{equation}

The predictions of the spectral tilt and tensor-to-scalar ratio for different values of $C_\Gamma$ and fixing the values of the mass scale $\mu=0.0001 M_4$, axion decay constant $f=0.02M_4$, $C=70$ and $M_5=7.5\times 10^{-4}M_4$ for 60 e-foldings in this case are summarized in table 2. The temperature of the thermal bath $(\tau)$ and dissipation parameter $Q$ calculated for different values of $C_\Gamma$ are also presented in the table. It is seen that the spectral index does not depend on $C_\Gamma$ in this case. The range of $C_\Gamma$ is found to be $[4.88\times10^{6}, 8\times10^{6}]$. The upper bound of the parameter $C_\Gamma$ is obtained from the condition $Q<0.1$ and the lower bound of the parameter $C_\Gamma$ is obtained from the condition of warm inflation $\tau>H$. The model is consistent with observational constraints on $n_s$ and $r$ in this case.

\begin{table}[h!]
 \caption{The values of the spectral index $n_s$, tensor to scalar ratio $r$, $Q$ and temperature of the thermal bath $\tau$ (in units of $M_4$) for different values of $C_\Gamma$ (in units of $M_4$) with $f=0.02M_4$, $\mu=0.0001 M_4$, $M_5=0.00075 M_4$ and $N=60$}
   
  \begin{center}
\begin{tabular}{lllll}
\hline\noalign{\smallskip}
 $C_\Gamma$ & $Q$  & $ n_s $  & $ r$   & $ \tau$ \\
 \noalign{\smallskip}\hline\noalign{\smallskip}
 % $4\times10^{6}$ & $0.0068$ & $0.962729$ & $1.91\times10^{-7}$ & $1.58\times10^{-6}$  \\

 % $4.87\times10^{6}$ & $0.0149$ & $0.962729$ & $1.906\times10^{-7}$ & $1.924\times10^{-6}$  \\
 
 $4.88\times10^{6}$ & $0.015$ & $0.962729$ & $1.906\times10^{-7}$ & $1.927\times10^{-6}$  \\
 
 $6\times10^{6}$ & $0.034$ & $0.962729$ & $1.903\times10^{-7}$ & $2.37\times10^{-6}$ \\

$8\times10^{6}$ & $0.1$ & $0.962729$ & $1.9\times10^{-7}$ & $3.16\times10^{-6}$ \\

% $1\times10^{7}$ & $0.26$ & $0.962729$ & $1.899\times10^{-7}$ & $3.95\times10^{-6}$ \\

% $1.2\times10^{7}$ & $0.55$ & $0.962729$ & $1.89\times10^{-7}$ & $4.74\times10^{-6}$ \\
\noalign{\smallskip}\hline

\end{tabular}
\end{center}
\end{table}

\begin{figure}[h!]
 \centering
 \begin{subfigure}[h]{0.49\textwidth}
     \centering
     \includegraphics[width=\textwidth]{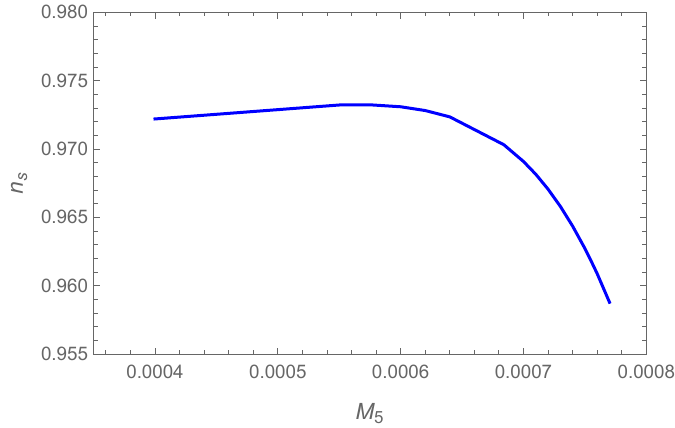}
    
 \end{subfigure}
 \hfill
 \begin{subfigure}[h]{0.49\textwidth}
     \centering
     \includegraphics[width=\textwidth]{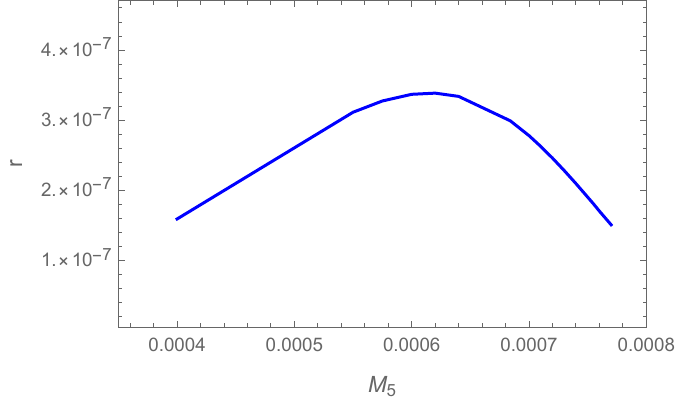}
     
 \end{subfigure}
 \caption{  Spectral index $n_s$ and tensor-to-scalar  ratio $r$ as a function of $M_5$ (in units of $M_4$) for $C_\Gamma=6\times 10^{6}$.}
    \label{f2}
 \end{figure} 
Fig.~\ref{f2} shows the variation of spectral index $n_s$ and tensor-to-scalar ratio $r$ as a function of $M_5$ (in units of $M_4$) for  $C_\Gamma=6\times 10^{6}$. From this plot, The range of $M_5$ (in units of $M_4$) is found to be $0.0007<M_5<0.00076$ for $C_\Gamma=6\times 10^{6}$.

\subsection{Numerical results high energy ($V\gg \lambda$) and strong dissipation ($Q\gg 1$) regime}
\subsubsection{Case-1: $\Gamma=\Gamma_0$}
The slow-roll parameters in this case are 
\begin{eqnarray}
    \epsilon=\frac{3 M_5^3}{4 \pi \Gamma_0 f^2}\frac { \left(1-\cos \left(\frac{\phi }{f}\right)\right)}{  \left(1+\cos \left(\frac{\phi }{f}\right)\right)}
\end{eqnarray}
\begin{eqnarray}
   \eta=-\frac{3 M_5^3}{4 \pi \Gamma_0 f^2}\frac { \cos \left(\frac{\phi}{f}\right)}{  \left(1+\cos \left(\frac{\phi }{f}\right)\right)} 
\end{eqnarray}
 In this case, the condition $\epsilon(\phi_{f})=1$ gives the final field value $\phi_f$ which is given by
 \begin{eqnarray}
     \phi_f= f\cos ^{-1}\left(\frac{3M_5^3-4\pi\Gamma_0 f^2 }{3M_5^3+4\pi\Gamma_0 f^2 }\right)
 \end{eqnarray}
  The number of e-foldings (equation \eqref{115}) in this case becomes, 
  \begin{eqnarray}
N =- \frac{4\pi}{3M_5^3}\int _{\phi_i}^{\phi_f}\frac{V\Gamma}{V^\prime}d \phi
\label{eee115}
\end{eqnarray}
The initial value of the inflaton field $(\phi_i)$ can be obtained by solving equation \eqref{eee115} and is given by
\begin{equation}
    \phi_i=2 f \sin ^{-1}\left(\sin \left(\frac{\phi_f}{2 f}\right) e^{-\frac{3 M_5^3N}{8 \pi \Gamma_0 f^2}}\right)
\end{equation}
Since, dissipation coefficient $\Gamma$ is independent of temperature in this case, so, there is no coupling between temperature and inflation field. Thus, the scalar power spectrum  $P_R$, the tensor power spectrum $P_T$, and the spectral index $n_s$ at initial field value $\phi_i$ become,
\begin{equation}
    P_R= \frac{8\pi^{1/4} f^{3/2}\mu^3 \Gamma_0^{9/4} \left(1+\cos\left(\frac{\phi_i}{f}\right)\right)^{9/4}}{3^{9/4} C^{1/4}   M_5^{27/4} \sin ^{\frac{3}{2}}\left(\frac{\phi_i}{f}\right)}
\end{equation}
\begin{equation}
    P_T=\frac{64\pi\left( \mu^4\left(1+\cos\left(\frac{\phi_i}{f}\right)\right)  \right)^2 \coth \left(\frac{k}{2 \tau  }\right)}{9 M_4^2 M_5^6}
\end{equation}
\begin{equation}
   n_s= 1-\left(\frac{27 M_5^3 \left(\mu ^4 \sin \left(\frac{\phi_i}{f}\right)\right)^2}{16 \pi f^2  \Gamma_0  \left( \mu^4\left(1+\cos\left(\frac{\phi_i}{f}\right)\right)  \right)^2}+\frac{9 M_5^3 \left(\mu ^4 \cos \left(\frac{ \phi_i}{f}\right)\right)}{  8 \pi f^2 \Gamma_0    \mu^4\left(1+\cos\left(\frac{\phi_i}{f}\right)\right)  }\right) 
\end{equation}

\begin{figure}[h!]
\centering
     \includegraphics[width=.6\textwidth ]{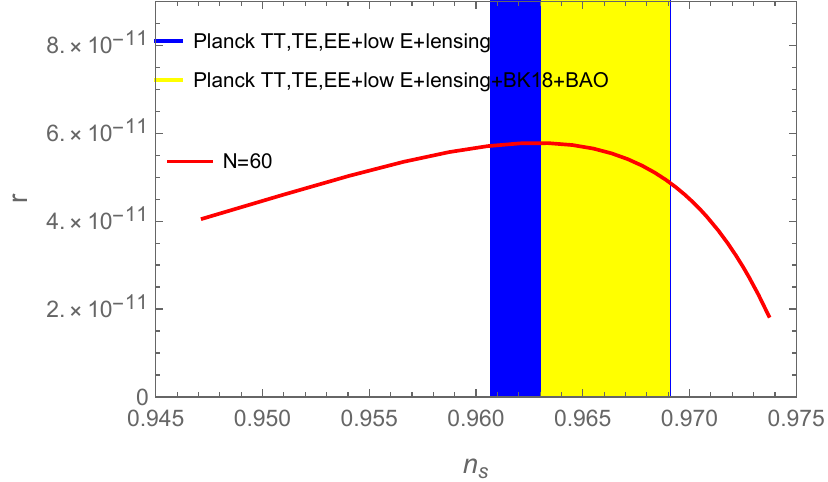}
      
 \caption{  The spectral index $n_s$, and the tensor-to-scalar ratio $r$ predicted by warm natural inflationary model in the braneworld scenario for $N=60$ in strong dissipative regime (red solid line). The marginalized joint $95\%$ CL regions for the spectral index $n_s$, and the tensor-to-scalar ratio $r$, obtained from Planck 2018 and lensing data alone, and their combinations with BICEP3/Keck Array (BK18) and BAO data are shown in blue and yellow respectively.}
    \label{f3}
\end{figure}

The predictions of the spectral tilt and tensor-to-scalar ratio in strong dissipation regime for this case are shown in  Figure~\ref{f3}. The plot is obtained by varying dissipation parameter $\Gamma_0$ and fixing the values of the mass scale $\mu=0.0001 M_4$, axion decay constant $f=0.001M_4$, $C=70$ and $M_5=6.8\times 10^{-4}M_4$ for 60 e-foldings. The range of parameter $\Gamma_0$ (in units of $M_4$) is found to be $[1.6\times 10^{-3}, 2.6\times 10^{-3}]$. The plot shows that the model is in good agreement with the Planck 2018 and BICEP3/Keck Array (BK18) data.

   \begin{figure}[h!]
 \centering
 \begin{subfigure}[h]{0.49\textwidth}
     \centering
     \includegraphics[width=\textwidth]{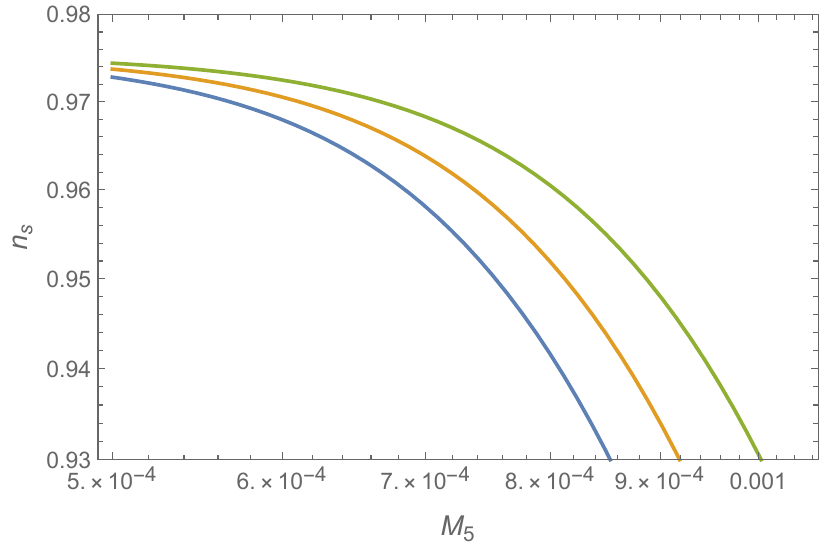}
    
 \end{subfigure}
 \hfill
 \begin{subfigure}[h]{0.49\textwidth}
     \centering
     \includegraphics[width=\textwidth]{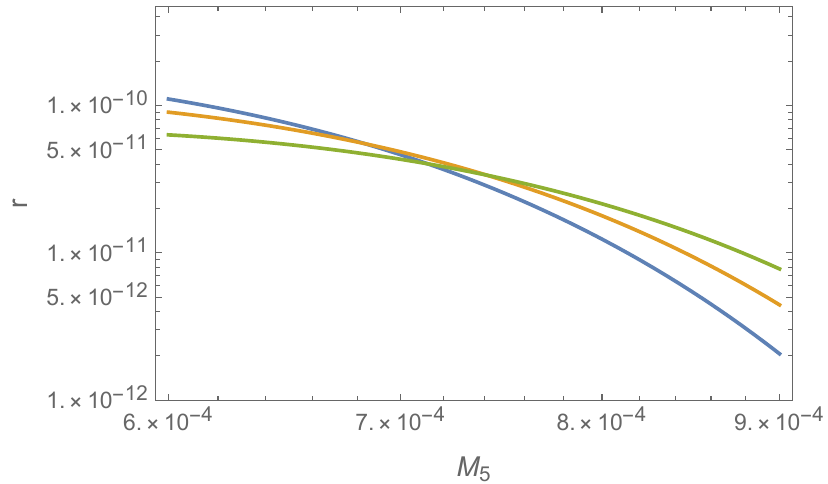}
     
 \end{subfigure}
 \caption{  Spectral index $n_s$ and tensor-to-scalar ratio $r$ as a function of $M_5$ (in units of $M_4$) for $\Gamma_0=1.6\times 10^{-3} M_4$ (blue), $ \Gamma_0=2\times 10^{-3} M_4 $ (orange) and $\Gamma_0=2.6\times 10^{-3} M_4$ (green).}
    \label{f4}
 \end{figure} 
 
Fig.~\ref{f4} shows the variation of spectral index $n_s$ and tensor-to-scalar ratio $r$ as a function of five-dimensional Planck mass $M_5$ (in units of $M_4$) for different values of the dissipation coefficient $\Gamma_0$. From this plot, we can obtain the allowed ranges of $M_5$.  For $\Gamma_0 = 1.6 \times 10^{-3} M_4$, the constraint on the five-dimensional Planck mass is found to be $0.000583 < M_5 < 0.000681$. For $\Gamma_0 = 2 \times 10^{-3} M_4$, the allowed range shifts to $0.000627 < M_5 < 0.000734$. In the case of $\Gamma_0 = 2.6 \times 10^{-3} M_4$, the constraint becomes $0.000678 < M_5 < 0.000799$. 

 \subsubsection{Case-2: $\Gamma(T)=C_\Gamma\frac{\tau^3}{f^{2}}$}
The slow-roll parameters in this case become
\begin{eqnarray}
    \epsilon=\frac{3 M_5^3}{4 \pi \Gamma f^2}\frac { \left(1-\cos \left(\frac{\phi }{f}\right)\right)}{  \left(1+\cos \left(\frac{\phi }{f}\right)\right)}
    \label{e46}
\end{eqnarray}
\begin{eqnarray}
   \eta=-\frac{3 M_5^3}{4 \pi \Gamma f^2}\frac { \cos \left(\frac{\phi}{f}\right)}{  \left(1+\cos \left(\frac{\phi }{f}\right)\right)}
   \label{e47}
\end{eqnarray}
and, when $Q\gg 1$, the equation of motion of the inflaton field becomes
\begin{equation}
 \dot{\phi}=-\frac{V^\prime}{\Gamma}
       \label{133}
 \end{equation}
 Equating $\tau=\left(\frac{\Gamma f^2 }{C_\Gamma}\right)^\frac{1}{3}$ (from $\Gamma=C_\Gamma\frac{\tau^3}{f^{2}}$) and $ \tau=\left(  \frac{\Gamma\dot{\phi}^2}{4HC}\right)^\frac{1}{4}$ (from equation \eqref{e20}), we get the expression for $\Gamma$ as
 \begin{equation}
  \Gamma=\frac{\left(\frac{3}{\pi }\right)^{3/7} C_\Gamma^{4/7} \mu ^{24/7} \sin ^{\frac{6}{7}}\left(\frac{\phi}{f}\right)}{2\ 2^{5/7} C^{3/7} f^2 \left(\frac{\mu ^4 \left(\cos \left(\frac{\phi}{f}\right)+1\right)}{M_5^3}\right)^{3/7}}
 \end{equation}
 Substituting the expression for $\Gamma$ in equations \eqref{e46} and \eqref{e47}, we obtain the slow-roll parameters $\epsilon$ and $\eta$ in terms of the model parameters. It is checked numerically that $\epsilon$ reaches unity faster, and thus, from equation $\epsilon=1$, the corresponding value of the inflaton field $(\phi_f)$ is obtained. The number of e-foldings (equation \eqref{115}) in this case becomes, 
  \begin{eqnarray}
N =- \frac{4\pi}{3M_5^3}\int _{\phi_i}^{\phi_f}\frac{V\Gamma}{V^\prime}d \phi
\label{e115}
\end{eqnarray}
Using the expression above of $N$, the initial field value $(\phi_i)$ can be obtained for 60 e-foldings.

In this case, the inflaton field is strongly coupled to the radiation field, and therefore, the factor $G(Q)$ will be present in the expression of the primordial power spectrum. Therefore, the expressions for the scalar power spectrum  $P_R$, and the tensor power spectrum $P_T$ at initial field value $\phi_i$ become, 
\begin{eqnarray}
 P_R&=& 
\frac{C_\Gamma^{\tfrac{9}{7}}\,\mu^{\tfrac{48}{7}}\,
\sin^{\tfrac{3}{7}}\!\left(\tfrac{\phi_i}{f}\right)\,
\left(1+\cos\!\left(\tfrac{\phi_i}{f}\right)\right)^{\tfrac{9}{7}}}
{3^{\tfrac{9}{7}}\,2^{\tfrac{6}{7}}\,\pi^{\tfrac{5}{7}}\,
C^{\tfrac{17}{14}}\,f^{3}\,M_5^{\tfrac{27}{7}}} G(Q)
\end{eqnarray}
\begin{equation}
    P_T=\frac{64\pi\left( \mu^4\left(1+\cos\left(\frac{\phi_i}{f}\right)\right)  \right)^2 \coth \left(\frac{k}{2 \tau  }\right)}{9 M_4^2 M_5^6}
\end{equation}
 From the above expressions of the power spectrum $P_R$ and $P_T$, the spectral index $n_s$ and tensor-to-scalar ratio $r$ are computed using equations \eqref{e16} and \eqref{e17} for different values of model parameters. 

\begin{figure}[h!]
\centering
     \includegraphics[width=.75\textwidth ]{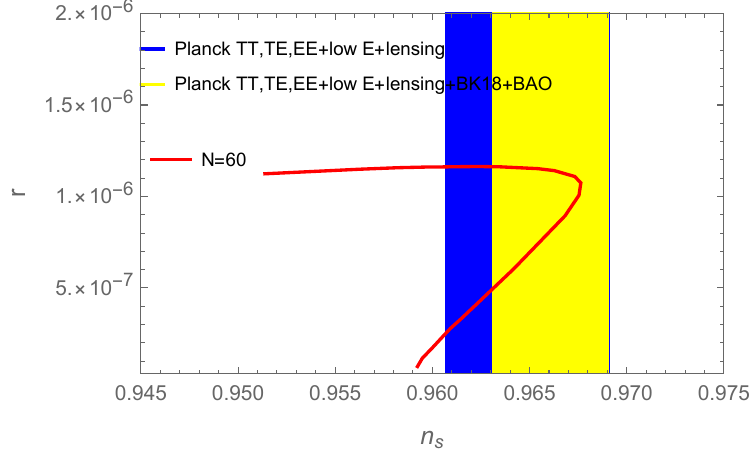}
      
 \caption{  The spectral index $n_s$, and the tensor-to-scalar ratio $r$ predicted by warm natural inflationary model in the braneworld scenario for $N=60$ in strong dissipative regime (red solid line). The marginalized joint $95\%$ CL regions for the spectral index $n_s$, and the tensor-to-scalar ratio $r$, obtained from Planck 2018 and lensing data alone, and their combinations with BICEP3/Keck Array (BK18) and BAO data are shown in blue and yellow respectively.}
    \label{f5}
\end{figure}

\begin{figure}[h!]
 \centering
 \begin{subfigure}[h]{0.49\textwidth}
     \centering
     \includegraphics[width=\textwidth]{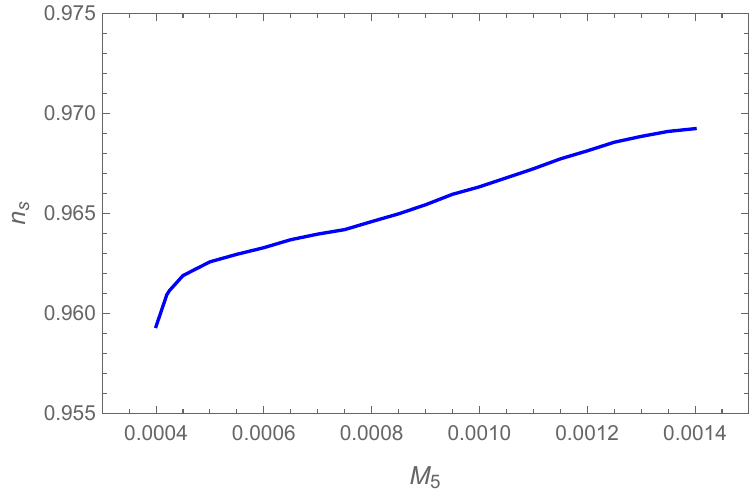}
    
 \end{subfigure}
 \hfill
 \begin{subfigure}[h]{0.49\textwidth}
     \centering
     \includegraphics[width=\textwidth]{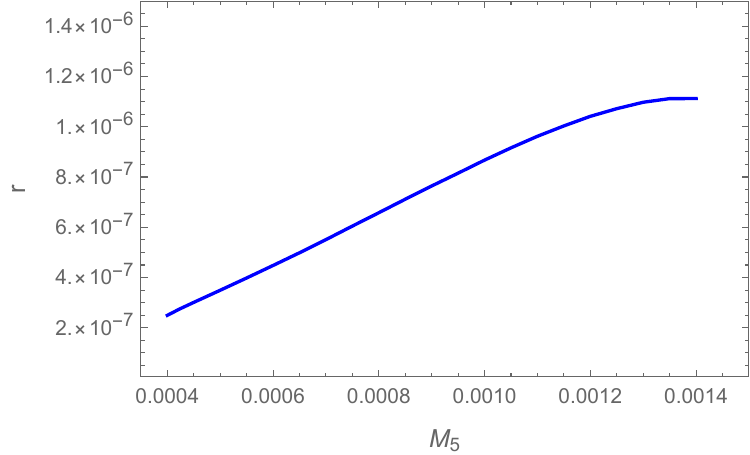}
     
 \end{subfigure}
 \caption{  Spectral index $n_s$ and tensor-to-scalar ratio $r$ as a function of $M_5$ (in units of $M_4$) for $C_\Gamma=2\times 10^{7} $}
    \label{f6}
 \end{figure}

The $n_s-r$ plot with the Planck 2018 and BICEP3 constraints is shown in Fig.~\ref{f5}. The plot is obtained by varying the parameter $C_\Gamma$ and fixing $\mu=0.0001 M_4$, $f=0.005M_4$, $C=70$ and $M_5=7.5\times 10^{-4}M_4$ for 60 e-foldings. The range of parameter $C_\Gamma$ is found to be $[5.25\times10^{6}, 1\times10^{8}]$ for which the values of $n_s$ and $r$ are in good agreement with Planck 2018 and BICEP data at $95\%$ CL. Fig.~\ref{f6} shows the variation of spectral index $n_s$ and tensor-to-scalar ratio $r$ as a function of $M_5$ (in units of $M_4$) for  $C_\Gamma=2\times 10^{7}$. From this plot, The range of $M_5$ (in units of $M_4$) is found to be $0.00042<M_5<0.00135$ for $C_\Gamma=2\times 10^{7}$.

 \vspace{.6cm}
 
 \section{Conclusion}
 In recent years, studying inflationary models in the context of higher-dimensional theories has become a significant research area.
 In this work, we analyzed a warm inflation model in braneworld cosmology with the natural potential in light of  Planck 2018 and BICEP data. Under the slow-roll approximations, we have derived equations that govern the background dynamics of this model in both strong and weak dissipation regimes. Considering two forms of dissipation coefficients: constant and temperature-dependent, we obtained expressions for scalar and tensor power spectra, scalar spectral index, and the tensor-to-scalar ratio in both regimes. We predicted the spectral tilt and the tensor-to-scalar ratio by varying $C_\Gamma$ for the variable dissipation case and varying $\Gamma_0$ for the constant diffipation case, while keeping the other parameters fixed. We also analyzed the dependence of the spectral index $n_s$ and the tensor-to-scalar ratio $r$ on the five-dimensional Planck mass $M_5$, and from which we obtained the allowed range of $M_5$ consistent with observational data.

 In weak dissipative regime, the range of $\Gamma_0$ (in units of $M_4$) is found to be $[9\times10^{-8}, 6\times10^{-7}]$ and the range of $C_\Gamma$ is found to be $[4.88\times10^{6}, 8\times10^{6}]$ for temperature dependent dissipation coefficient. It is found that for the allowed parameter range, the obtained values of $n_s$ and $r$ are well within the limits set by the Planck 2018 data and the joint Planck and BICEP results, thus confirming the viability of the model in the weak dissipative regime.
 
 In strong dissipative regime, the range of parameter $\Gamma_0$ (in units of $M_4$) is found to be $[1.6\times 10^{-3}, 2.6\times 10^{-3}]$ for constant dissipation coefficient and the range of $C_\Gamma$ is found to be $[5.25\times10^{6}, 1\times10^{8}]$ for temperature dependent dissipation coefficient. The obtained values of $n_s$ and $r$ are consistent with the observational constraints from the Planck 2018 data and the joint Planck and BICEP results. Thus, the model is viable in the strong dissipative regime also.

 In addition, the value of the decay constant of the axion $f$ is set well below the Planck scale, and in the strong dissipative regime, it is below the GUT scale. These lower values of $f$ make the warm inflationary model on brane theoretically more consistent.

\newpage
 \printbibliography
\end{document}